\documentclass[XeLaTex,twocolumn,epjc3]{svjour3}       
\RequirePackage[T1]{fontenc}

\smartqed  

\RequirePackage{graphicx}
\RequirePackage{mathptmx}      
\RequirePackage{flushend}
\usepackage{caption}
\usepackage{subcaption}
\usepackage{amsmath}
\usepackage{amssymb}
\usepackage{epstopdf}
\usepackage{float}
\RequirePackage[numbers,sort&compress]{natbib}
\RequirePackage[colorlinks,citecolor=blue,urlcolor=blue,linkcolor=blue]{hyperref}

\begin{document}

\title{Solutions for neutron stars in General Relativity from a complexity structure scalar boundary condition}

\author{Robert S. Bogadi \thanksref{addr1,e1}
       \and
             Megandhren Govender \thanksref{addr1,e2} 
        \and 
              Genly Leon \thanksref{addr2,addr3,e3}
        \and
            Andronikos Paliathanasis \thanksref{addr3,addr4,addr5,addr6,e4} 
}

\thankstext{e1}{e-mail: bogadi.robert@gmail.com}
\thankstext{e2}{e-mail: megandhreng@dut.ac.za}
\thankstext{e3}{e-mail: genly.leon@ucn.cl}
\thankstext{e4}{e-mail: anpaliat@phys.uoa.gr}

\institute{Department of Mathematics, Faculty of Applied Sciences, Durban University of Technology, Durban 4000, South Africa \label{addr1} 
\and
Departamento de Matem\'aticas, Universidad Cat\'olica del Norte, Avdenida Angamos 0610, Casilla 1280, Antofagasta, Chile, Antofagasta, Chile  \label{addr2} 
\and
Institute of Systems Science, Durban University of Technology, Durban 4000, South Africa \label{addr3}
\and
Centro de Investigaci\'on, Innovaci\'on y Creaci\'on (CIIC), Universidad
Cat\'olica de Temuco, Temuco, Chile \label{addr4}
\and
Departamento de Ciencias Matem\'{a}ticas y F\'{\i}sicas, Facultad de Ingenier\'{\i}a, Universidad Cat\'olica de Temuco, Temuco, Chile \label{addr5}
\and
National Institute for Theoretical and Computational Sciences (NITheCS), South Africa  \label{addr6}
}

\date{Received: date / Accepted: date}

\maketitle

\begin{abstract}
A new condition involving the complexity factor, a structure scalar arising from the orthogonal splitting of the Riemann tensor, is generated at the boundary of a spherically symmetric anisotropic fluid. This condition leads to the generation of a model, without invoking any further constraints on geometry or matter variables. The model generated is physically reasonable and suggests that there is a non-zero, positive minimum constraint on the complexity scalar value evaluated at the surface. This suggests that vanishing complexity is a globalized constraint which does not apply at the boundary, specifically for compact objects in the strong-gravity regime.
\end{abstract}


\section{Introduction}

The modelling of neutron stars which are the most compact, visible objects in the Universe are studied theoretically using Einstein's theory of general relativity (GR) \cite{yak} and the numerous types of extensions known as modified gravity theories which have progressed in sophistication \cite{sta,lob}. The application of relativistic gravity theory to dense, compact matter has even invoked higher dimensional frameworks for investigating neutron star stability, structure and properties \cite{maz}. Neutron stars have been subdivided into categories which include pulsars, accreting neutron stars and magnetars amongst others \cite{bor}, and also the hypothesized quark and strange-quark stars in search of the true ground state of matter \cite{zha}. The densities of such objects may exceed nuclear saturation density, thus particularly relevant to the formation, reactions and stability of fundamental elementary particles. 

Theoretical modelling of neutron stars using GR was pioneered by Schwarzschild, Tolman, Oppenheimer and Volkoff \cite{schw,tol,opp}. The Schwarzschild interior solution was the first exact solution for a perfect fluid with constant density. Tolman obtained solutions for spheres with variable density, a requirement for avoiding non-causality in the speed of sound. Oppenheimer and Volkoff investigated the equilibrium of cold, neutron-star matter which lead to the Tolman-Oppenheimer-Volkoff (TOV) equations for relativistic hydrostatic equilibrium \cite{tol,opp}. The TOV equations in combination with an equation of state (EoS) then provide a working model for a compact object such as a neutron star. Analytical equations of state range from a simple linear EoS with extension to include the MIT Bag constant \cite{azi}, to more complex ones such as quark-parametrized EoS's \cite{roch}. Since there are numerous EoS candidates available \cite{ji}, an all encompassing model is still out of reach. This is further compounded by the postulation of neutron star variants, namely, quark, hybrid, hyperon and nucleon stars.   

In developing a model via Einstein's theory of relativity, closure of the system of equations is required and is often achieved by specifying an EoS, however, it is of greater interest and value to solve the system by some other means and then analyse the resulting equation of state. This is made possible by specifying a geometric constraint rather than a physical one. A physical constraint such as pressure isotropy or imposing an EoS might ensure physical viability but it is after all, ideal, and results in a toy model or a purely mathematical result. 

The rich geometrical aspect of GR allows for the study of curvature and structure scalar invariants. This includes the method of embedding four-dimensional spacetime into higher dimensional manifolds \cite{kar} and the formulation of gravity theories from Lagrangians (modified gravity) \cite{lov}, of which GR is a consequence, has yielded successful models of compact objects. 

A recent addition to modelling compact objects is the complexity factor as presented by Herrera and co-workers \cite{her1,her2}. It is a structure scalar invariant arising from the orthogonal splitting of the Riemann tensor. It should also be noted that a previous definition of complexity involving statistical measure has been used in the past \cite{ave}. Herrera's definition is more applicable to astrophysical bodies as it is linked to the Riemann curvature tensor. Numerous models have been constructed by imposing a vanishing complexity constraint, both in GR and in modified gravitational theories \cite{ari,nas,mau}. Thus we see its versatility in obtaining solutions from gravity field equations. \\
In this study, we solve a differential equation involving the complexity factor for the non-vanishing case. Anisotropic models are then generated by stipulating an ansatz for the temporal potential in the line element and setting the complexity factor parameter at the surface boundary.

\section{Interior geometry}

Modelling of a star in four-dimensional spacetime, in the absence of shearing stresses, is commonly done using a spherically symmetric line element, 

\begin{equation} \label{1} 
ds^2 = -A(r)^2 dt^2 + B(r)^2[dr^2 + r^2 (d \theta^2 + \sin^2 \theta d \phi^2)]\,,
\end{equation} 

in which the metric functions $g_{00} = A(r)^2$ and $g_{11} = B(r)^2$, describing the gravitational potentials, are yet to be determined. This shear-free metric in isotropic coordinates has been used to obtain exact solutions in General Relativity and extended gravity theories, and has been successfully applied to both self-gravitating static and dynamic, collapsing systems \cite{kra,stew,bon,gov}. The energy momentum tensor for an anisotropic fluid is given by 

\begin{equation} \label{2} 
T_{ab} = diag\left(-\rho~,~p_r~,~p_t~,~p_t \right) ,
\end{equation} 

where $\rho$, $p_r$ and $p_t$ are the energy density, radial and transverse pressures of the fluid respectively. The fluid four-velocity ${\bf u}$ is comoving and given by

\begin{equation}
u^a = \frac{1}{A}\delta_0^a~.
\end{equation}

The Einstein field equations \{$R_{ab} - \frac{1}{2}R g_{ab} = \frac{8\pi G}{c^4} T_{ab}$\} describing the interior of the stellar fluid then become

\begin{eqnarray}   \label{efe}
8\pi \rho_0 &=&  - \frac{1}{B^2}\left(2\frac{B''}{B} - \frac{{B'}^2}{B^2} + \frac{4}{r}\frac{B'}{B} \right),  \label{efe1} \\ 
8\pi p_{r0} &=& \frac{1}{B^2} \left[\frac{{B'}^2}{B^2} + 2\frac{A'}{A}\frac{B'}{B} + \frac{2}{r}\left(\frac{A'}{A} + \frac{B'}{B}\right) \right],  \label{efe2}  \\ 
8\pi p_{t0} &=& \frac{1}{B^2}\left[\frac{A''}{A} + \frac{B''}{B}  - \frac{{B'}^2}{B^2} +  \frac{1}{r}\left(\frac{A'}{A} + \frac{B'}{B}\right)\right], \label{efe3}  
\end{eqnarray}

where geometric units have been used ($G = c = 1$). The interior line element (\ref{1}) must be matched to the exterior Schwarzschild solution in comoving isotropic coordinates \cite{bon}, given by

\begin{equation}
ds^2 = -\left(\frac{1 - \frac{M}{2r}}{1 + \frac{M}{2r}}\right)^2 dt^2 + \left(1 + \frac{M}{2r}\right)^4 \left(dr^2 + r^2d\Omega^2\right)
\end{equation}

where $M$ is the mass within a sphere of radius $R$ and $r \geq R$. The mass may be computed using the interior solution, and is given by

\begin{eqnarray}
M &=& \Bigg[ - r^2 B' - \frac{r^3 B'^2}{2 B} \Bigg]_{\Sigma}, \label{mass}
\end{eqnarray} 

where $\Sigma$ represents the boundary surface between the interior and exterior spacetimes such that $R = r_\Sigma B(r_\Sigma)$.

We note that the shear-free approximation is prone to instability and that pressure anisotropy $\delta=p_r-p_t$, in some respects, can assist in overcoming this problem \cite{her1}. In our study and application to modelling, we evaluate the adiabatic index and the sound speeds in order to investigate stability and physical viability.

\section{The complexity structure scalar~ $\it{Y_{TF}}$}

The complexity factor for a general, self-gravitating shear-free system follows from Herrera's definition \cite{her3} and is given by

\begin{equation}\label{YTF_gen}
Y_{TF} = 8\pi \delta + \frac{4\pi}{(r B)^3}\int_0^r (r B)^3 \left( \rho ' - 3q B \frac{\dot{B}}{A}\right) dr~.
\end{equation}

We note that this is a general definition in so far as it accommodates gravitational collapse with heat flow within a shear-free system.\\
By substituting the associated Einstein field equations \cite{bog2}, (\ref{YTF_gen}) becomes

\begin{eqnarray} \label{YTF_int}
Y_{TF} &=& 8\pi \delta - \frac{1}{(r B)^3}\int_0^r 2r\bigg[r \left(B'' + \frac{r}{2} B''' \right) \nonumber \\
&& - \frac{r B'}{B} \left(3 B' +2 r B'' \right)  - B'  + \frac{r^2 (B')^3}{B^2} \bigg]  dr 
\end{eqnarray}

which can be integrated to yield

\begin{equation} \label{YTF}
Y_{TF} = 8\pi \delta - \frac{1}{B^2}\left[\frac{B ''}{B} - 2\left(\frac{B '}{B}\right)^2 - \frac{1}{r}\frac{B '}{B} \right]
\end{equation} 
as was shown by Bogadi et al. \cite{bog2}. This is readily solved for systems with vanishing complexity for both the isotropic and anisotropic cases. \\

We now extend this to include systems for which the complexity factor is non-vanishing ($Y_{TF} \neq 0$). For the most general case which includes anisotropy, ($\delta \neq 0$), the complexity factor (\ref{YTF}) may be written,

\begin{equation} \label{YTF_A}
Y_{TF} = \frac{1}{B^2}\left[ \frac{1}{A}\left(A '' - \frac{A '}{r} \right) - 2\frac{A '}{A}\frac{B '}{B}\right]
\end{equation}

which is still in general, both spatially and temporally dependent. In considering a solution for $B(r,t)$, we note that (\ref{YTF_A}) is not quite Riccati, and follows the form
\[ B' = f(r,t)B + g(r,t)Y_{TF}B^3 \]
with $f$ and $g$ being functions of $A(r,t)$. \\

For vanishing complexity, it was shown that one easily obtains the condition $A '  = C(t) B^2 r$ which reduces the problem of finding solutions to a single-generating function \cite{bog2}. In the case of non-vanishing complexity, integration of (\ref{YTF_A}) is more challenging. A first approach is to consider equation (\ref{YTF_A}) at the surface boundary and set $(A'/B)_\Sigma = \alpha$ where $\Sigma$ represents the boundary. This leads to the boundary complexity condition,

\begin{equation}
\alpha ' - \left(\frac{B'}{B} + \frac{1}{r}\right)\alpha = Y_{TF}|_\Sigma A B
\end{equation} 
which can be partly integrated to give us the expression,

\begin{equation}
A' = B^2 r \left(C(t) + \int \frac{A}{r}Y_{TF} dr \right).
\end{equation}
Clearly one can calculate the metric potential $B(r,t)$ for a known potential $A(r,t)$ according to

\begin{equation}
B = \sqrt{\frac{A'/r}{\left(\int \frac{A}{r}Y_{TF} dr + C(t) \right)}}
\end{equation}
provided the integration in the denominator can be performed.

Again this reduces to the expression obtained previously for vanishing complexity factor. In order to generate analytical solutions, one should choose suitable metric functions for $A(r,t)$ such that the required integration can be performed. Since $Y_{TF}$ is considered at the surface boundary, it can be taken outside of integration. \\

Another approach that does not rely on the integrability of $A(r,t)$ proceeds as follows. We reconsider (\ref{YTF_A}) in the form,

\begin{equation}\label{YTF_A2}
B' - \frac{1}{2} \left(\frac{A''}{A'} - \frac{1}{r}\right)B + \frac{Y_{TF}}{2\alpha}A B^2 = 0
\end{equation} 

in which we have considered only partial conversion of $A'/B$ to $\alpha$, and a constant complexity factor $Y_{TF}$. This can now be integrated to give

\begin{equation} \label{anisol_Y}
B = C(t) \sqrt{\frac{A'}{r}} \exp{\left(-\frac{Y_{TF}}{4\alpha^2}A^2\right)}.
\end{equation}

Again this is in agreement with the expression obtained for vanishing complexity and satisfies the original equation (\ref{YTF_A}). This consideration allow us to investigate the deviation of the solutions when a for a nonzero complexity factor. 

We note that the factor $Y_{TF}/4\alpha^2$ in the exponent functions as a single parameter and make the assumption that the metric potentials maintain their spatial and temporal dependence although (\ref{YTF_A}) is strictly speaking a boundary condition. This assumption has been used numerous times by researchers in considering the temporal boundary condition for radiating, collapsing systems $(p_r = q B)|_\Sigma$ \cite{san} in which the same parameter $\alpha = (A'/B)_\Sigma$ completely determines the temporal behaviour for the entire system \cite{bon,iva,pin,tew,pal,das}. Similarly, expression (\ref{anisol_Y}) functions as a boundary condition, providing a link between the gravitational potentials.

\section{Application to modelling}

We now consider a simple static model, firstly choosing an appropriate ansatz. Previously, the ansatzes have been applied to the spatial part $B$, but in our case we must specify metric potential $A_0(r)$. In the context of massive, compact matter such as a neutron star, we choose the ansatz

\begin{equation} \label{A0}
A_0(r) = a\left(1 + k r^2 \right)^\kappa.
\end{equation}
 
The later consideration ansatz (\ref{A0}) follows the Durgapal framework \cite{dur} and is also a simplified form of a solved metric potential obtained by Govender and Thirukkanesh \cite{gov} in isotropic coordinates. We see that if $\kappa = 1$, we obtain a conformally flat spacetime which would be more suited to cosmological systems. For compact objects such as neutron stars, suitable models are obtainable for $\kappa \leq 0.5$. 
 
Considering (\ref{anisol_Y}) in the static regime, we establish

\begin{equation} \label{B0}
B_0(r) = b\sqrt{2a~k~\kappa\left(1 + k r^2 \right)^{\kappa - 1}}~ \exp{\left(- y~A_0(r)^2\right)}
\end{equation}
where $y = \frac{Y_{TF}}{4\alpha^2}$ and $b$ is a constant. \\
\\

Matching of the interior metric to an exterior Schwarzschild metric in isotropic coordinates is given according to

\begin{eqnarray}
A_0(\Sigma) &=& \frac{1 - \frac{M}{2R}}{1 + \frac{M}{2R}} \\
B_0(\Sigma) &=& \left(1 + \frac{M}{2R}\right)^2
\end{eqnarray}

where $\Sigma$ is the boundary coordinate such that $B_0(\Sigma) \Sigma = R$, the radius of the star.
we obtain the expressions for constants $a$ and $b$, namely

\begin{eqnarray}
a &=& \left(\frac{1 - \mu}{1 + \mu}\right) \left(1 + \frac{k R^2}{\left(1 + \mu\right)^4}\right)^{-\kappa} \\
b &=& \sqrt{\frac{1}{2\kappa}\left(\frac{1 + \mu}{1 - \mu}\right)\left(R^2 + \frac{1}{k}\left(1 + \mu\right)^4\right)}\times \nonumber \\
&& \exp{\left(y\left(\frac{1 - \mu}{1 + \mu}\right)^2\right)}
\end{eqnarray}

where $\mu = \frac{M}{2R}$. By applying the boundary condition for the radial pressure ($p_r0(\Sigma) = 0$), we obtain an expression for the curvature parameter $k$,

\begin{equation}
k = \frac{\left(1 + \mu\right)^4}{R^2} \left(\frac{\mu}{\nu}\sqrt{1 - 4\mu} - 1\right) 
\end{equation}

where $\nu = \mu\left(1 - 5\mu\right) + \left(1 - 4\mu\right)\frac{R^2}{4}Y_{TF}$. \\

The mass function is given by

\begin{equation}
m_0(r) = -r^2 B_0'(r) - \frac{r^3 B_0'(r)^2}{2B_0(r)}.
\end{equation}

The mass of the star ($M = m_0(\Sigma)$) is then used to obtain an expression for the index parameter $\kappa$. We calculate

\begin{equation}
\kappa = \frac{\mu^2}{\left(\mu\sqrt{1 - 4\mu} - \nu\right)}
\end{equation}

The metric functions are now specified in terms of parameters $\{M,R,Y_{TF}\}$. We write (\ref{A0}) and (\ref{B0}) compactly as

\begin{equation}
A_0(r) = \frac{1 - \mu}{1 + \mu}\left(\frac{1 + k r^2}{1 + k\left(\frac{R}{(1 + \mu)^2}\right)^2}\right)^{\frac{\mu^2}{\left(\mu\sqrt{1 - 4\mu} - \nu\right)}}
\end{equation}

and

\begin{eqnarray}
B_0(r) &=& (1+\mu)^2\left(\frac{1 + k r^2}{1 + k\left(\frac{R}{(1 + \mu)^2}\right)^2}\right)^{\frac{1}{2}(\frac{\mu^2}{\left(\mu\sqrt{1 - 4\mu} - \nu\right)}-1)} \nonumber \\
&&\exp\bigg[\frac{-1}{4\mu^2}\left(\mu(1 - 5\mu) - \nu\right)\times \nonumber \\
&&\left(1 - \left(\frac{1 + k r^2}{1 + k\left(\frac{R}{(1 + \mu)^2}\right)^2}\right)^{\frac{2\mu^2}{\left(\mu\sqrt{1 - 4\mu} - \nu\right)}}\right)\bigg].
\end{eqnarray}

The surface parameter $\alpha = (A'/B)|_\Sigma$ is also calculated to be

\begin{equation}
\alpha = \frac{M}{R^2\sqrt{1 - \frac{2M}{R}}} \left(\frac{1 - \frac{M}{2R}}{1 + \frac{M}{2R}}\right) = g_s \left(\frac{1 - \frac{M}{2R}}{1 + \frac{M}{2R}}\right)
\end{equation}

where $g_s$ is the surface gravity. Thus we see that $\alpha$ is closer to the surface gravity than in previous scenarios where $\alpha = M/R^2$ \cite{bog3}. This would affect the temporal progression of a dynamical self-gravitating system. \\  

We investigate the suitability of these potentials in describing the compact object Cen X-3 with parameters $M = 1.49~M_\odot$ and $R = 9.178~ km$. These values have less uncertainty than many others derived from observational data. Setting the complexity structure factor at $Y_{TF} = 0.0005~ km^{-2}$, comparable to values for such a parameter \cite{bog}, we obtain plots of the matter variables as shown in Figures 1 and 2. Further quantities are calculated in order to access stability, namely the sound speed profiles and adiabatic index. These are shown in Figures 3 and 4. Expressions for the sound speeds are given in appendix B.

\begin{figure}
\centering
\includegraphics[scale=0.85]{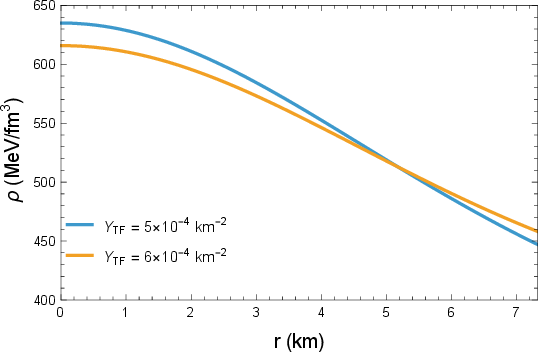}\caption{Density profile for Cen~X-3}
\end{figure}

\begin{figure}
\centering
\includegraphics[scale=0.85]{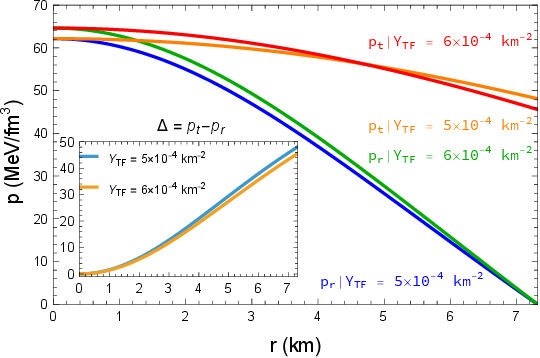}\caption{Radial and tangential pressure profiles for Cen~X-3}
\end{figure}

\begin{figure}
\centering
\includegraphics[scale=0.85]{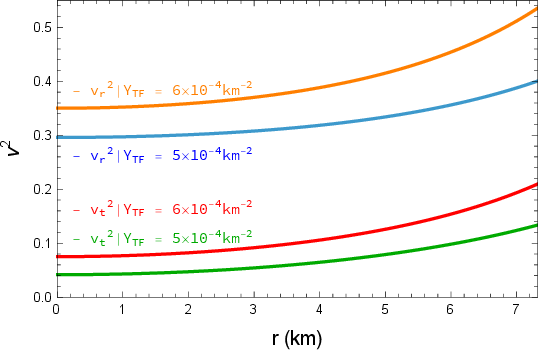}\caption{Sound speed squared for Cen~X-3}
\end{figure}

\begin{figure}
\centering
\includegraphics[scale=0.85]{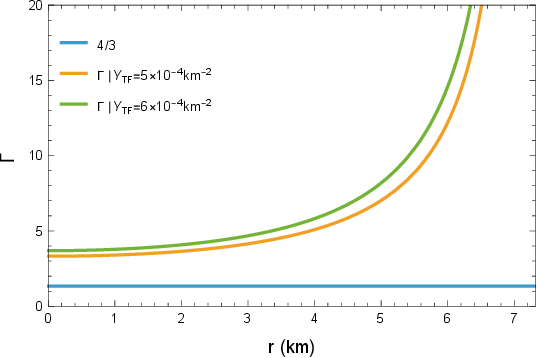}\caption{Adiabatic index for Cen~X-3}
\end{figure}

\begin{figure*}
\centering
\begin{subfigure}{.5\textwidth}
  \centering
  \includegraphics[scale=0.8]{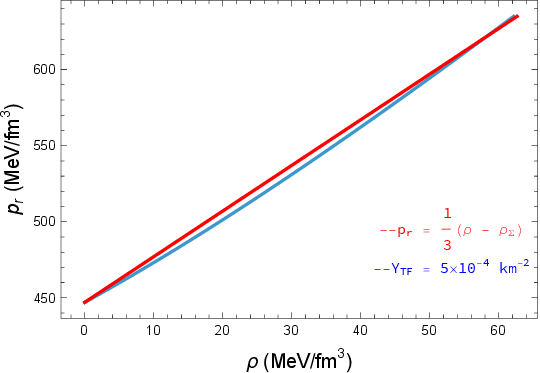}\caption{$Y_{TF}|_\Sigma = 5\times10^{-4}km^{-2}$}
\end{subfigure}%
\begin{subfigure}{.5\textwidth}
  \centering
  \includegraphics[scale=0.8]{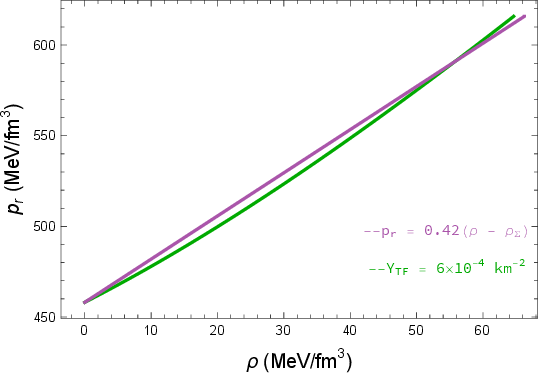}\caption{$Y_{TF}|_\Sigma = 6\times10^{-4}km^{-2}$}
\end{subfigure}
\caption{Model EoS compared with linear EoS for Cen X-3}
\end{figure*}

Model parameters for a list of well studied stars are given in Table 1 (Appendix A). Curvature parameter $k$ may be compared with coefficients of other models which contain the $(1 + k r^2)$ factor. Surface and core densities, and central pressures are given in Table 2 (Appendix A). 
Mass radius curves have been calculated (Figure 5) by considering an average surface density of $\rho(\Sigma) = 8.0\times 10^{14} g/cm^3$ for the selected stars (Appendix A). 

\begin{figure}
\centering
\includegraphics[scale=0.85]{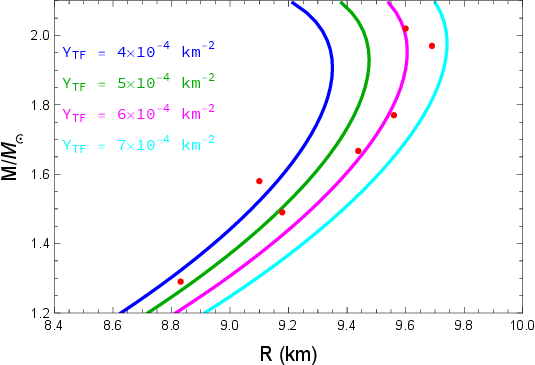}\caption{Mass-radius curves generated for $\rho(\Sigma) = 8.0\times 10^{14} g/cm^3$~;~ Red dots denote selected stars (Appendix A)}
\end{figure}

\section{Discussion}

The plots of the physical quantities are reasonable for the complexity values used. The energy density and pressure profiles (Figures 1 and 2) show expected trends and the magnitudes of quantities are reasonable, also shown in Table 2 for other stars. The sound speed profiles (Figure 3) are physically reasonable, showing a monotonic increase towards the surface of the star. Similar behaviour was obtained for a Finch and Skea stellar model \cite{shar} and a Durgapal IV model with minimal geometric deformation \cite{ort}. We note that values of the complexity factor less than $4\times 10^{-4} km^{-2}$ are unfavourable as the tangential sound speed squared profile approaches zero and can become negative which is not only physically unviable, but mathematically undesirable as the sound speed would become a complex quantity. The tangential sound speed is less than the radial sound speed which is a requirement for stability \cite{rat}. Furthermore, the adiabatic index (Figure 4) is greater than the Chandrasekhar limit of $4/3$, thus promoting stability. The mass-radius curves show that most stars may be accommodated using the model. 

Constant surface density was used to generate the curves since the zero pressure boundary condition, being a function of the mass and radius, does not naturally lead to the required parametric plotting with variation of a curvature parameter which is itself a function of mass and radius. Additional curves can be generated for specific densities aimed at candidate stars, with varying complexity factors. \\
A main result is that the value of the complexity factor at the boundary is non-zero and comparable to results obtained by Bogadi et al. \cite{bog}. Their static model employed a Vaidya-Tikekar potential and a linear equation of state (EoS). No EoS was assumed in our study. We plot the resulting EoS of our model in Figure 5 and compare this with an approximating linear EoS.

\section{Conclusion}

We have shown the feasibility of constructing a model of a compact object by specifying the complexity structure scalar at the surface boundary of the object. Neither pressure isotropy nor an equation of state was imposed, in contrast to the usual practice; the only geometric input is the ansatz (\ref{A0}) for the temporal potential.
The solution obtained for the metric function $B(r)$ is implicit in as far as the surface parameter $\alpha$ having been calculated using both metric functions. The model has an intrinsic, implicit and thus hidden nature which could be applied to machine and deep learning strategies in future work. Indeed, the surface parameter $\alpha$ is also used for gravitational collapse and dictates the temporal behaviour of the process.\\
A related approach has been presented recently within gravitational decoupling according to the minimal geometric deformation method in \cite{car,zub}, where the complexity factor is
used as the auxiliary condition required to fix the deformation function. In \cite{car,zub}, polynomial complexity factors are considered for the reconstruction of gravitational models. Physically accepted solutions were found to lead to constraints for the coefficients of the complexity factor. In this work we followed a different approach. We chose isotropic coordinates and worked within the bounds of General Relativity (GR), without introducing an additional source or a decoupling parameter so that the complexity condition alone closes the system. Moreover, we imposed the complexity constraint only at the surface and let the interior profile to be derived from the gravitational theory of GR.\\

Two limitations appear to be evident: Firstly, the framework was established within the class of models generated by the Durgapal ansatz (\ref{A0}), and the extent to which the bounds are ansatz-independent is yet to be determined; secondly, the treatment is static -- extending the construction to the radiating case is natural, given the role played by $\alpha$ in both cases. 
Finally, the differential equation (\ref{YTF_A}) for the complexity structure scalar is non-trivial, and more general solutions might be obtained using Lie symmetry analysis \cite{pal2}.

\appendix

\onecolumn

\section{Model parameters and calculated matter quantities for selected stars} \label{A1}

\begin{table}[H]
\begin{tabular}{|c|c|c|c|c|c|}
\hline
{\it Star} & $ M/M_\odot $ & $ R $ & $ k $ & $\kappa$ & $\alpha$   \\
    &   & $(km)$ & $(km^{-2})$ &   & $(km^{-1})$   \\
\hline
Cen~X-3 & 1.49~$\pm$~0.08 & 9.178~$\pm$~0.13 & 0.0115 & 0.436 & 0.0285     \\
Vela~X-1 & 1.77~$\pm$~0.08 & 9.56~$\pm$~0.08 & 0.0164 & 0.429 & 0.0323     \\
PSR~J1614-2230 & 1.97~$\pm$~0.04 & 9.69~$\pm$~0.2 & 0.0233 & 0.427 & 0.0362    \\
PSR~J1903+0327 & 1.667~$\pm$~0.021 & 9.438~$\pm$~0.03 & 0.0143 & 0.431 & 0.0307   \\ 
LMC~X-4 & 1.29~$\pm$~0.05 & 8.831~$\pm$~0.09 & 0.00912 & 0.446 & 0.0261   \\
4U~1820-30 & 1.58~$\pm$~0.06 & 9.1~$\pm$~0.4  & 0.0147 & 0.428 & 0.0312  \\
4U~1636-536 & 2.02~$\pm$~0.12 & 9.6~$\pm$~0.6 & 0.0280 & 0.427 & 0.0385   \\
\hline
\end{tabular}
\centering \caption{\label{Tabel-1} Neutron star data \cite{azi} and model parameters for ($Y_{TF}|_\Sigma = 5\times10^{-4}km^{-2}$)}
\end{table}

\begin{table}[H]
\begin{tabular}{|c|c|c|c|c|c|}
\hline
{\it Star} & $ M/M_\odot $ & $ R $ & $\rho_c$ & $\rho_s$ & $p_{rc}$   \\
    &   & $(km)$ & $\times10^{15}(g/cm^3)$ & $\times10^{15}(g/cm^3)$ & $\times10^{35}(dyne/cm^2)$   \\
    &   &             &  $Y_{TF}^{(5)}~~~|~~~Y_{TF}^{(6)}$ & $Y_{TF}^{(5)}~~~|~~~Y_{TF}^{(6)}$ & $Y_{TF}^{(5)}~~~|~~~Y_{TF}^{(6)}$  \\
\hline
Cen~X-3 & 1.49~$\pm$~0.08 & 9.178~$\pm$~0.13 & 1.13~~~|~~~1.10 & 0.797~~~|~~~0.817 & 0.996~~~|~~~1.04     \\
Vela~X-1 & 1.77~$\pm$~0.08 & 9.56~$\pm$~0.08 & 1.32~~~|~~~1.29 & 0.778~~~|~~~0.797 & 1.35~~~|~~~1.39     \\
PSR~J1614-2230 & 1.97~$\pm$~0.04 & 9.69~$\pm$~0.2 & 1.58~~~|~~~1.54 & 0.769~~~|~~~0.788 & 1.81~~~|~~~1.86    \\
PSR~J1903+0327 & 1.667~$\pm$~0.021 & 9.438~$\pm$~0.03 & 1.24~~~|~~~1.20 & 0.784~~~|~~~0.804 & 1.20~~~|~~~1.24   \\ 
LMC~X-4 & 1.29~$\pm$~0.05 & 8.831~$\pm$~0.09 & 1.03~~~|~~~1.00 & 0.809~~~|~~~0.829 & 0.821~~~|~~~0.860   \\
4U~1820-30 & 1.58~$\pm$~0.06 & 9.1~$\pm$~0.4  & 1.31~~~|~~~1.27 & 0.831~~~|~~~0.851 & 1.21~~~|~~~1.25  \\
4U~1636-536 & 2.02~$\pm$~0.12 & 9.6~$\pm$~0.6 & 1.76~~~|~~~1.71 & 0.780~~~|~~~0.799 & 2.11~~~|~~~2.15   \\
\hline
\end{tabular}
\centering \caption{\label{Tabel-1} Values of physical quantities computed for $Y_{TF}^{(5)} = 5\times10^{-4}km^{-2}$ and $Y_{TF}^{(6)} = 6\times10^{-4}km^{-2}$}
\end{table}

\section{Expressions for radial and tangential sound speed squared} \label{B1}

\begin{eqnarray}
v_r^2 &=& - \bigg(\mu\sqrt{1 - 4\mu} \left(\mu^2(\mu^2 - (1 - 4\mu) - 2\nu(4 - k r^2)) - 3\nu^2 \right)  \nonumber \\
&& + \mu^2\left(4\mu^2(1 - 4\mu) + \nu(4\nu + 3(1 - 4\mu) - \mu^2) +  (\mu^2(5\mu^2 - (1 - 4\mu)) - \nu^2)k r^2 \right) + \nu^3  \nonumber \\
&& + \left(2\mu(\mu\nu - \sqrt{1 - 4\nu} (\mu^2 - \nu k r^2)) - (\mu^2(1 - 4\mu + \mu^2) + \nu^2)k r^2 \right)(\mu(1 - 5\mu) - \nu)\left(\frac{1 + k r^2}{1 + \frac{k R^2}{(1 + \mu)^4}}\right)^{2\kappa} \nonumber \\
&& + \left(\mu\sqrt{1 - 4\mu} - \nu + 3\mu^2 k r^2 \right)(\mu(1 - 5\mu) - \nu)^2 \left(\frac{1 + k r^2}{1 + \frac{k R^2}{(1 + \mu)^4}}\right)^{4\kappa} + (\mu(1 - 5\mu) - \nu)^3 k r^2 \left(\frac{1 + k r^2}{1 + \frac{k R^2}{(1 + \mu)^4}}\right)^{6\kappa} \bigg) / D  \nonumber \\
\end{eqnarray}

\begin{eqnarray}
v_t^2 &=& -2 \bigg(\mu\sqrt{1 - 4\mu}\left(\mu^2(\mu^2 - (1 - 4\mu) - 4\nu) - 3\nu^2 \right) + 2\mu^2\left(\mu^2(1 - 4\mu + \mu^2 k r^2) + \nu^2\right) \nonumber \\
&& - \nu\left(\mu^2(\mu^2 - 3(1 - 4\mu)) - \nu^2 \right) + \left(\mu\sqrt{1 - 4\mu} - \nu + 2\mu^2 k r^2\right)(\mu(1 - 5\mu) - \nu)^2 \left(\frac{1 + k r^2}{1 + \frac{k R^2}{(1 + \mu)^4}}\right)^{4\kappa} \bigg)/D
\end{eqnarray}

where 

\begin{eqnarray}
D &=& \mu\sqrt{1 - 4\mu}\left(5\mu^2(\mu^2 - (1 - 4\mu)) - \nu(15\nu - 2\mu^2k r^2)\right)  +  \mu^2(\mu^2(\mu^2 - (1 - 4\mu)) - \nu^2)k r^2 + 5\nu(\mu^2(3(1 - 4\mu) - \mu^2) + \nu^2) \nonumber \\
&& - (2\mu(5\mu(\mu\sqrt{1 - 4\mu} - \nu) - \nu\sqrt{1 - 4\mu} k r^2) + (\mu^2(1 - 4\mu + 9\mu^2) + \nu^2) k r^2)(\mu(1 - 5\mu) - \nu)\left(\frac{1 + k r^2}{1 + \frac{k R^2}{(1 + \mu)^4}}\right)^{2\kappa} \nonumber \\
&& + (5(\mu\sqrt{1 - 4\mu} - \nu) - 7\mu^2 k r^2)(\mu(1 - 5\mu) - \nu)^2\left(\frac{1 + k r^2}{1 + \frac{k R^2}{(1 + \mu)^4}}\right)^{4\kappa} + (\mu(1 - 5\mu) - \nu)^3 k r^2 \left(\frac{1 + k r^2}{1 + \frac{k R^2}{(1 + \mu)^4}}\right)^{6\kappa}  \nonumber 
\end{eqnarray}

\twocolumn

\paragraph{Data Availability Statement:} Data sharing not applicable–no new data generated, the article describes
entirely theoretical research.

\paragraph{Acknowledgements:} RB and MG acknowledge support from the office of the Deputy Vice-Chancellor for Research and Innovation at the Durban University of Technology. GL and AP were partially supported by FONDECYT Grant 1240514, ETAPA 2026. The authors acknowledge the COST Action CA23130 ``Bridging high and low energies in search of quantum gravity (BridgeQG)''.


\end{document}